 \documentclass[reprint,amsmath,amssymb,aps,pre]{revtex4-1}

\usepackage{amsmath,amssymb,amsfonts,amstext,graphics,graphicx,subfigure,dcolumn,bm}
\usepackage{setspace}
\usepackage{mathtools}
\usepackage[colorlinks]{hyperref}
\usepackage{color}

\usepackage{mathbbol}
\usepackage{natbib}

\def\XXint#1#2#3{{\setbox0=\hbox{$#1{#2#3}{\int}$ }
\vcenter{\hbox{$#2#3$ }}\kern-.5\wd0}}

\def\calb{\mathcal{B}}

\def\calj{\mathcal{J}}

\def\call{\mathcal{L}}

\def\bfm{\mathbf{m}}

\def\bfxi{\mathbf{\xi}}

\def\R{\mathbb{R}}

\def\bq{\begin{equation}}
\def\eq{\end{equation}}
\def\bqy{\begin{eqnarray}}
\def\eqy{\end{eqnarray}}

\def\bal#1\eal{\begin{align}#1\end{align}}

\def\quadd{\quad\quad}

\def\al{\alpha}
\def\be{\beta}

\def\de{\delta}

\def\ep{\epsilon}

\def\et{\eta}
\def\ga{\gamma}
\def\Ga{\Gamma}

\def\ka{\kappa}
\def\la{\lambda}
\def\La{\Lambda}
\def\na{\nabla}

\def\Om{\Omega}

\def\rh{\rho}
\def\si{\sigma}
\def\Si{\Sigma}
\def\ta{\tau}

\def\ze{\zeta}

\def\bfJ{\mathbf{J}}

\def\bfx{\mathbf{x}}

\def\bfZ{\mathbf{Z}}
\def\bfv{\mathbf{v}}
\def\bfw{\mathbf{w}}
\def\bfp{\mathbf{p}}

\def\bfz{\mathbf{z}}

\def\bfq{\mathbf{q}}

  \def\bfxi{\boldsymbol{\xi}}
    \def\bfze{\boldsymbol{\ze}}

\def\p{\partial}
\def\na{\nabla}
\def\andq{\quad\mathrm{and}\quad}
\def\andqq{\quadd\mathrm{and}\quadd}

\def\nn{\nonumber}

\def\bq{\begin{equation}}
\def\eq{\end{equation}}
\def\bqy{\begin{eqnarray}}
\def\eqy{\end{eqnarray}}

\def\bal#1\eal{\begin{align}#1\end{align}}

\def\4tensLa{\bar{\bar{\Lambda}}}

\def\tensD{\bar{D}}
\def\tenska{\bar{\kappa}}
\def\tensJ{\bar{J}}

\def\al{\alpha}
\def\be{\beta}

\def\de{\delta}

\def\ep{\epsilon}

\def\et{\eta}
\def\ga{\gamma}
\def\Ga{\Gamma}

\def\ka{\kappa}
\def\la{\lambda}
\def\La{\Lambda}
\def\na{\nabla}

\def\Om{\Omega}

\def\rh{\rho}
\def\si{\sigma}
\def\Si{\Sigma}
\def\ta{\tau}

\def\ze{\zeta}

\def\p{\partial}
\def\na{\nabla}
\def\andq{\quad\mathrm{and}\quad}
\def\andqq{\quadd\mathrm{and}\quadd}

\def\nn{\nonumber}

\begin{document}

\title{
Metriplectic 4-bracket algorithm for constructing thermodynamically consistent dynamical systems}
\author{Azeddine Zaidni}
\email{azeddine.zaidni@um6p.ma}
\affiliation{Mohammed VI Polytechnic University, College of Computing, Lot 660, Hay Moulay Rachid, Ben Guerir, 43150, Morocco}
\author{Philip J.~Morrison}
\email{morrison@physics.utexas.edu}
\affiliation{Department of Physics and Institute for Fusion Studies, 
The University of Texas at Austin, Austin, TX, 78712, USA}

\date{\today}

\begin{abstract}

A unified thermodynamic  algorithm is presented for constructing  thermodynamically consistent dynamical systems, i.e., systems that have Hamiltonian and dissipative parts that conserve energy while producing entropy.  The algorithm is based on the metriplectic 4-bracket given in  Morrison and Updike  [Phys.\ Rev.\ E {\bf 109}, 045202 (2024)].  A feature of the unified thermodynamic  algorithm is the force-flux relation 
$\mathbf{J}^\alpha =  - L^{\alpha\beta}\,  \nabla(\delta H / \delta \xi^\be)$ for phenomenological coefficients $L^{\al\be}$, Hamiltonian $H$ and dynamical variables $\xi^\be$.  The algorithm is applied to the Navier-Stokes-Fourier, the Cahn-Hilliard-Navier-Stokes,    and  Brenner-Navier-Stokes-Fourier systems, and significant generalizations  of these systems are obtained.

\end{abstract}

\maketitle

\tableofcontents

\section{Introduction}

 Metriplectic dynamics was  established in the 1980s  \cite{pjm84,pjm84b,pjm86} to provide a  framework for describing joined Hamiltonian and dissipative dynamics with the property  that  thermodynamic consistency is guaranteed.  (See \cite{pjmK82,kaufman1984dissipative,pjmH84,grmela1984particle} for different attempts at incorporating dissipation in a framework.)   Thermodynamic consistency means the joined Hamiltonian and dissipative system  conserves energy, consistent with the first law of thermodynamics, and produces entropy,   consistent with the second law.

In 1997 the name GENERIC was proposed  \cite{ottinger1, grmela} for a framework that is equivalent to metriplectic dynamics (see page 11 of \cite{pjmU24}).  In a sequence of works \cite{ottinger1,grmela, ottinger2005} these authors were the first to explicitly  incorporate  ideas from  non-equilibrium thermodynamics  (e.g.\  \cite{de2013non}) into the framework.  Specifically,   Onsager's  reciprocal relations  \cite{Onsager,Casimir} were employed to  ensure entropy production.  More recently, the same connection between non-equilibrium thermodynamics theory and metriplectic dynamics  was  made in \cite{coquinot2020general} for a general class of magnetofluid models and more generally  in  \cite{pjmU24} where the metriplectic 4-bracket, a convenient quantity  for constructing  thermodynamically consistent systems, was introduced.

In the theory of non-equilibrium thermodynamics, it is assumed that the fluxes, say $\bfJ^\al$,  are typically linear functions of thermodynamic forces (sometimes called {\it affinities}), say $X_\be$;  i.e., 
\bq
\bfJ^\al =   L^{\al\be} X_\be\,,
\label{onsager1}
\eq
 where $L^{\al\be}$ is a symmetric matrix, $\al$ and $\beta$ are indices  for the set of  dynamical variables, and the repeated $\be$ index is to be summed.

In our previous work \cite{zaidni2}, we proposed an algorithm for constructing a metriplectic 4-bracket and, consequently,    a means for  producing  thermodynamically consistent systems. This was done for a general  Navier-Stokes Cahn-Hilliard system, a  model for two-phase flow. The algorithm has the following four steps: i) First, select a set of dynamical variables. ii) Next, select energy and entropy functionals, $H$ and $S$, dependent on the dynamical variables, based on the physics of the phenomena to be described. iii) The third step is to obtain the noncanonical Poisson bracket \cite{pjm98} of the ideal (non-dissipative) part of the theory, with the chosen entropy as a Casimir invariant. iv) The final step is to construct a metriplectic 4-bracket. We will refer to the algorithm as the ``unified thermodynamic algorithm'' or UT algorithm for short.

In  previous works \cite{pjmU24,zaidni2,pjmS24}  step iv) relied on past experience and intuition.  This  final step of the UT algorithm was  facilitated by making use of  the Kulkarni-Nomizu product \cite{Kulkarni, nomizu1971spaces} (K-N product) of two operators $M$  and $\Si$. This shifts  the burden to determining the two operators, which still might not be straightforward.   The question of how to determine the  $M$  and $\Si$ leads us to express the dissipative fluxes in a  new manner, one  different from the Onsager reciprocal relations approach of \eqref{onsager1}.   Instead we assume a flux-force relation as follows:
\bq
\bfJ^\al =  - L^{\al\be}\,  \na(\delta H / \delta \xi^\be)\,, 
\label{genOnsag}
\eq
where  $\xi^\be$ are the dynamical variables as known from step i) of the UT algorithm,  $H$ is the Hamiltonian as obtained from step ii), and $\delta H / \delta \xi^\be$ is the functional derivative.  We will see that expression \eqref{genOnsag} is intimately connected to the distinctive physical roles played by  $M$  and $\Si$,  and it guides their determination.   Expression \eqref{genOnsag} can be further generalized (see \eqref{Flux-VF} below) by replacing $\nabla$ by any pseudodifferential operator that has an adjoint, instead of a simple spatial gradient. This adjoint assumption is crucial for developing a method to derive the metriplectic 4-bracket and complete the UT algorithm.  

Consequently, the algorithm can proceed using the inputs: the Hamiltonian functional $H$, the entropy functional $S$,  and the unknown coefficients $L^{\al\be}$. These coefficients are determined by assuming phenomenological laws, as is done with the forces  in non-equilibrium thermodynamics theory \cite{ottinger2005}.  This approach also has the added feature that it makes clear the origin of  dependencies on dynamical and  thermodynamic variables; viz.,  the forces that arise from $\delta H / \delta \xi^\be$, which depends on  internal energy functionals through $H$,  and  those that arise from phenomenological laws through $L^{\al\be}$.

This paper is structured as follows: In Sec.~\ref{Overview on metriplectic framework}, we provide an overview of the metriplectic framework, where foundational concepts are reviewed, starting with the Hamiltonian formalism (Sec.~\ref{Hamiltonian formalism}) and moving to the metriplectic 4-bracket formalism (Sec.~\ref{Metriplectic 4-bracket formalism}). At the end of this section we discuss the critical features needed in order to develop a systematic method to build the metriplectic 4-bracket.  In  Sec.~\ref{Derivation of metriplectic 4-bracket}, we focus on the derivation of the metriplectic 4-bracket. Here, a systematic approach to the theory is developed (Sec.~\ref{K-N Construction and Fluxes}), followed by a discussion on the relationship to  non-equilibrium thermodynamics principles (Sec.~\ref{non-equilibrium thermodynamics theory}). In  Sec.~\ref{Examples}, we present three examples to illustrate the application of the developed theory by applying  the UT algorithm. Specifically, the Navier-Stokes-Fourier (NSF) system (Sec.~\ref{NSF}), the Cahn-Hilliard-Navier-Stokes (CHNS) system (Sec.~\ref{CHNS}), and the Brenner-Navier-Stokes-Fourier (BNSF) system (Sec.~\ref{BNSF}) are explored as cases demonstrating the theory’s flexibility and general applicability. We observe that the  NSF and BNSF systems are special cases of a general theory we develop.  Finally,  in Sec.~\ref{conclusion} we briefly summarize and make a few comments about ongoing and future work.

\section{Overview on metriplectic framework}
 \label{Overview on metriplectic framework}
 
\subsection{Hamiltonian formalism}
\label{Hamiltonian formalism}

Let us briefly recall the Hamiltonian formalism in infinite dimensions. The first step of the UT algorithm   presented in \cite{zaidni2} involves selecting a set of dynamical variables. It is preferable to choose conserved quantities as the variables. For example, in fluid dynamics, one might select the mass density, momentum density, and entropy density. In  general, we consider the dynamics of classical field theories involving multi-component fields
\bq
\bfxi(\bfz, t)=\left(\xi^1(\bfz, t), \xi^2(\bfz, t), \ldots, \xi^N(\bfz, t)\right)
\label{xi}
\eq 
defined on $\bfz=(z^1,z^2,\dots, z^n)\in \Om$ for times $t \in \mathbb{R}$. Here we use $\bfz$ to be a label space coordinate with the volume element $ d^n\!z$, but with the domain $\Om$ unspecified. For example, in fluid mechanics $\Om$  would be the 3-dimensional domain occupied by the fluid   and we will use  $\bfx=(x^1,x^2,x^3)$ to indicate a point in $\Omega$ for this case. In general we suppose that $\xi^1,\ldots,\xi^N$ can be real-valued scalars or densities defined on  space-time $\Om\times\mathbb{R}$, vector fields in the tangent or cotangent bundles of the manifold $\Om$, or even elements in its tensor bundle.  Thus, for some $\al$, $\xi^\al$ could be a scalar, a vector, or any tensorial quantity that  is convenient for the system being described.  We will forgo formal geometric considerations and suppose our infinite-dimensional phase space has coordinates $\bfxi=(\xi^1,\ldots,\xi^N)$ and observables are  functionals  that map  $\bfxi\mapsto \R$ at each fixed time.  We will denote the space of such functionals by $\calb$.  Then a Poisson  bracket is  an antisymmetric bilinear operator
\bq
\{\,\cdot\,,\,\cdot\,\}\colon  \calb \times \calb  \rightarrow \calb\,,
\eq
where this bracket is assumed to satisfy the Jacobi identity, 
$\{\{F, G\}, K\} + \{\{K, F\}, G\} + \{\{G, K\}, F\} = 0$, thereby providing a realization of a Lie algebra [see, e.g., \cite{Sudarshan} chap.\ 14],  and in addition fulfill the Leibniz rule. A general infinite-dimensional  form of this bracket,  for any given two functionals $F,G\in\calb$, can be  written as follows:
\bq
\{\,F, G \,\} =  \int_\Om \! \!  d^n\!z \!\ \int_\Om \! \! d^n\!z' \,  \calj^{\al\be}   \frac{\delta F}{\delta \xi^\al(z)} \frac{\delta G}{\delta \xi^\be\left(z'\right)} \,, 
\label{1b}
\eq
where  $\calj^{\al\be}(z,z')$ is a 2-tensor functional operator, that is antisymmetric, with coordinate form given by the following integral kernel:
\bal
\calj^{\al\be}(z,z')[\bfxi]=\calj(\mathbf{d}\xi^\al(z), \mathbf{d}\xi^\be(z'))[\bfxi]\, ,
 \nn
\eal
where $\al,\be$ range over $1,2,\dots, N$, and $\de F/\de\xi^{\al}$, $\de G/\de\xi^{\be}$ are the functional derivatives (see, e.g., \cite{pjm98} for a formal review of these notions). 
 
Upon inserting any functional of $\bfxi$, say an observable $\xi^\al$,  into the Poisson bracket  its evolution is determined by 
\bq
    \partial_t \xi^\al= \{\xi^\al, {H}\}\,, 
    \label{fH}
\eq
where $H[\bfxi]\in\calb$ is a  Hamiltonian functional. Here and henceforth we use the shorthand $\p_t =\p/ \p t$ for the partial derivative with respect time and we  will use an overdot to mean the total derivative $d/dt$, i.e,  $\dot{F} = dF/dt$.  For example the evolution of the Hamiltonian functional is given by $\dot{H} = \{H,H\} = 0$, due to the antisymmetry of the bracket. We will also use the shorthand $\p_i =\p/ \p z^i$ for the partial derivative with respect to the spatial variable $z_i$. Casimir invariants are special functionals $\mathfrak{C}$ that satisfy $ 
\{F, \mathfrak{C}\}= 0$  for any functional $F$, and thus are constants of motion for any Hamiltonian. 

The second step of the UT algorithm is the selection of the Hamiltonian functional and a Casimir invariant to serve as the entropy. The choice of these functionals is based on the physics of  the phenomena one wishes to describe.  However, across all the cases we have examined, the Hamiltonian functional is the total energy of the system  and the usual total entropy of the system is a Casimir invariant. 

The construction of the noncanonical Poisson bracket \eqref{1b} is the third step of UT algorithm.  Since the publication of  \cite{pjmG80}, there is a huge literature on this for a variety of systems, e.g., \cite{pjm82,pjm98,pjmT00,hamdi,pjmDL16,pjmC20, pjmU24,pjmS24} give  Poisson brackets for a great many systems, including fluid dynamics, magneto-fluid dynamics, two-phase fluid dynamics, plasma kinetic theory and so one.

\subsection{Metriplectic 4-bracket formalism}
\label{Metriplectic 4-bracket formalism}

Step iv), the final step of the UT algorithm,  is the construction of the metriplectic 4-bracket. This construction was  introduced in \cite{pjmU24} to describe   dissipative dynamics. We briefly recall the metriplectic 4-bracket description for  infinite-dimensional systems. In this description, we consider the dynamics of classical field theories with multi-component fields as presented in \eqref{xi}.   We define the 4-bracket on functionals as  
\bq
(\,\cdot\,,\,\cdot\,\,;\,\cdot\,,\,\cdot\,)\colon \calb \times \calb \times \calb \times \calb \rightarrow \calb\,,
\eq
 such that for any four functionals $F,K,G,N\in\calb$ we have 
\bal
(F, K ; G, N)&=  \int_\Om  \! \!  d^n\!z \!\! \int_\Om \! \! d^n\!z'\,\! \! \int_\Om  \! \!  d^n\!z'' \!\! \int_\Om \! \! d^n\!z'''\,  \hat{R}^{\al\be\ga\de}
\nn \\
& \times \frac{\delta F}{\delta \xi^\al(z)} \frac{\delta K}{\delta \xi^\be\left(z'\right)} \frac{\delta G}{\delta \xi^\ga\left(z''\right)} \frac{\delta N}{\delta \xi^\de\left(z'''\right)}\,, 
\label{2b}
\eal
where  $\hat{R}^{\al\be\ga\de}(z,z',z'',z''')$ is a 4-tensor functional operator with coordinate form given by the following integral kernel:
\bal
 \hat{R}^{\al\be\ga\de}(z,z',& z'',z''')[\bfxi]\nn\\
 &=\hat{R}(\mathbf{d}\xi^\al(z), \mathbf{d}\xi^\be(z'),
 \mathbf{d} \xi^\ga(z''), \mathbf{d} \xi^\de(z''))[\bfxi]\, ,
 \nn
\eal
where $\al,\be, \ga,\de$ range over $1,2,\dots, N$.
The 4-bracket is assumed to  satisfy the following proprieties:\\
(i) Linearity in all arguments, e.g, for all $\la\in\R$
\bq
(F + \la H, K; G, N) = (F, K; G, N) + \la (H, K; G, N)
\label{2B1}
\eq
(ii) The algebraic  symmetries
\bal (F, K; G, N) &= -(K, F; G, N)
\label{2B2}\\
(F, K; G, N) &= -(F, K; N, G) 
\label{2B3}\\
(F, K; G, N) &= (G, N; F, K)
\label{2B4} 
\eal
(iii) Derivation in all arguments, e.g.,
\bq(F K;G, N) = F(H, K; G, N) + (F, K; G, N)H\,.
\label{2B6}
\eq
Here, as usual, $FH$ denotes point-wise multiplication.

One way to create a specific metriplectic 4-bracket  that has  the requisite symmetry  properties \eqref{2B1}-\eqref{2B6} is by using the Kulkarni-Nomizu (K-N) product \citep{Kulkarni,nomizu1971spaces}. (See also \cite{fiedler03} for relevant theorems.)  Given two symmetric operator fields, say $\Si$ and $M$, the K-N product is defined as follows: 
\bal
  (\Sigma \wedge M)\left(dF,dK, dG,dN\right) 
&= \,\Sigma\left(dF, dG \right) M\left(dK, dN\right)
\nn\\
&\hspace{-.5cm} -\Sigma\left(dF , dN \right) M\left(dK , dG \right) \nn\\
&\hspace{-.5cm}+M\left(dF , dG \right) \Sigma\left(dK , dN \right)
\nn\\
&\hspace{-.5cm} -M\left(dF , dN \right) \Sigma\left(dK , dG \right)\,.
\eal
Thus, consistent with the bracket formulation of \eqref{2b}, we define a 4-bracket according to
\bq
(F, K ; G, N)=\int_\Om \! d ^n z\,W\,(\Sigma \wedge M)\left(dF, dK, dG, dN\right)\label{3b}
\eq
where $W$ is an arbitrary weight function, depending on $\bfxi$ and $z$, that multiplies $\Sigma \wedge M$. For the general forms of the bilinear operators $\Si$ and $M$, we refer to \citep{pjmU24, zaidni2};  here, we omit the details for brevity.

The 4-bracket tool plays a crucial role in the dissipative description of dynamics, provided it satisfies certain properties that guarantee the thermodynamic consistency -- namely, the first law (energy conservation) and the second law (entropy production). These properties are referred to 
as ``\textit{minimal metriplectic properties} ''.

Let $H$ be the Hamiltonian functional associated to the Poisson bracket \eqref{1b} and $S$ its Casimir invariant. As we mentioned previously, In the vast majority of infinite-dimensional dynamics, particularly in fluid dynamics, $H$ and $S$ present the total energy and the total entropy, respectively. Thus, the minimal metriplectic properties are the combination  of the requisite symmetries \eqref{2B1}-\eqref{2B6} and the positive semi-definiteness in the following manner: The sectional curvature defined as $K(H,S):=(S, H; S, H)$ should be non-negative
\bq
K(H,S)\geq 0\,.
\label{seccom}
\eq
The 4-brackets arrising  from a K-N product \eqref{3b}, will have  the minimal metriplectic proprieties if both $\Si$ and $M$ are positive. If one of $\Si$ or $M$  is positive definite,  defining an inner product, then the sectional curvature satisfies $(S,H;S,H)\geq 0$ with equality if and only if  $\de S/\de \bfxi \propto \de H/\de \bfxi$. The proofs of these results were first established in \citep{pjmU24} for the finite-dimensional case, and later extended to the infinite-dimensional case in \citep{zaidni2}.

Now, for any observable functional of  $\bfxi$, say $\xi^\al$,  its dissipative evolution is prescribed by
\bq
\partial_t {\xi^\al} =  (\xi^\al, H; S, H)\,.
\eq
Thus we have thermodynamic consistency because
\bal
\dot{H}&=\left(H, H ; S, H\right)\equiv 0\,,
\label{dotH1}
\\
\dot{S}&=\left(S, H ; S, H\right)= K(H,S) \geq 0\,,
\label{dotS1}
\eal
where \eqref{dotH1} follows from  the antisymmetry condition of \eqref{2B2} and  \eqref{dotS1} follows  from \eqref{seccom}, i.e., that   the sectional curvature being  non-negative.

We remind the reader, that the 4-bracket automatically gives  metriplectic   2-brackets (see \cite{pjmU24} for discussion), such as those in the early works \cite{pjm84,pjm84b,pjm86} via $(F,G)_H=(F,H;G,H)$. Because of the symmetries of the 4-bracket, $\dot{H}$ will vanish for any $H$, as opposed to designing the earlier 2-brackets to make this happen for specific Hamiltonians.

In our previous work \citep{zaidni2},  the fourth and final step of the UT algorithm, was achieved by constructing  the 4-bracket via the K-N product using the following  general forms for  the bilinear operators $\Si$ and $M$:   
\bal
M(dF ,dG ) &=  F_{\xi^{\al}}A^{\al\be}G_{{\xi^{\be}}}\,,
\label{M}
\\
\Si (dF ,dG ) &=  
\call^{(\al)} (F_{\xi^\al})\cdot B^{\alpha\beta} \! \cdot \call^{(\be)} (G_{\xi^\be}) 
\label{Sgma}\,,
\eal
where we compactified our notation by defining $F_{\xi^{\al}}\!\!\coloneqq\de F/\de \xi^{\al}$ and $F_{\bfxi}\coloneqq \left({ F}_{\xi^1},{F}_{ \xi^2},\ldots,{F}_{\xi^N}\right)$. Here, the repeated indices are to be summed, $A^{\al\be}$ and $B^{\alpha\beta}$ are symmetric in  $\al,\be=1,\dots,N$, i.e., $A^{\al\be}=A^{\be\al}$ and $B^{\al\be}=B^{\be\al}$,  
and $\call^{(\al)}$, for $\al\in \{1,\ldots,N\}$, is contained within a general class of pseudodifferential operators on $\calb$.  We have placed parenthesis around the upper index of  $\call^{(\al)}$  to emphasize that this index  is not to be summed. When a specific value is placed in this slot  there is no confusion and the parentheses will be dropped.  We will see later that the parameters $A^{\al\be}$ and $B^{\al\be}$ could be scalars, 2-tensors, 3-tensors or even  4-tensors, depending on the tensorial character of $\xi^\al$ and the type of dissipation phenomena. The $``\cdot"$ of \eqref{Sgma} then symbolizes  the appropriate contractions.

This final step of the UT algorithm, as implemented  in  the previous works  \citep{pjmU24, pjmS24,zaidni2},   has  two  avenues  for  criticism.   First, in the  examples of previous works, in particular for the  NSF and CHNS systems, the definitions of $M$ and $\Si$ were not established in a systematic or methodological way.  Rather, they were engineered to give desired results.  Second, since we aim to develop a general dissipative dynamics formalism, independently of  specification of the thermodynamics, it is reasonable that  quantities should depend only on the selected set of dynamical variables, with the exception of coefficients of  phenomenological laws. For example, in  previous works, various  factors of  $1/T$  were inserted in various places in an  ad hoc manner.  Is this inserted  temperature  determined by  the internal energy function of $H$ or is it some other  phenomenological assumption?

In  Sec.~\ref{K-N Construction and Fluxes}, we will propose an unambiguous  method for  choosing  the operators $M$ and $\Si$, and thereby overcoming these criticisms by a direct  construction of the metriplectic 4-bracket.  This approach is general and applicable to a broad range of infinite-dimensional systems. Various types of fluid dynamics, magnetofluid dynamics, two-phase fluid flows,  and so one, are particular cases.  We will see our construction can significantly generalize systems in the literature.

\section{Derivation of metriplectic 4-brackets}
\label{Derivation of metriplectic 4-bracket}

\subsection{Systematic development of the theory}
\label{K-N Construction and Fluxes}

In this section we give our  method for constructing the metriplectic 4-bracket.  Thus, as discussed  in Sec.~\ref{Metriplectic 4-bracket formalism},   the UT algorithm becomes complete if we accomplish the  final step by selecting the bilinear symmetric operators $M$ and $\Si$ of the K-N product.  We provide a direct procedure for making these selections. 

En route to our goal, we make some notational choices.   In the first step of the UT algorithm, the selected  set of dynamical variables defined on space-time $\Om\times\mathbb{R}$ was defined as  follows: 
\bq
\bfxi(\bfz, t)=\left(\xi^1(\bfz, t), \xi^2(\bfz, t), \ldots, \xi^N(\bfz, t)\right)\,,
\label{xi_x}
\eq
where we previously commented that it is preferable to choose the $\xi^\al$ to be densities. To be more specific, here we suppose 
 $\xi^1(\bfz, t), \xi^2(\bfz, t), \ldots, \xi^{N-1}(\bfz, t)$ satisfy conservation laws  and the last component, $\xi^N$,  represents the entropy density, i.e, the entropy per unit volume.  In practice the various $\xi^\al$ besides the entropy $\xi^N$  may, based on the physical properties under consideration,  have particular tensorial qualities, e.g., they may be scalars, vectors, or tensors or pseudo-tensors of arbitrary rank.   To avoid a clutter of notation, we will not be explicit about this tensorial character, but strive for a notation that makes it clear how to proceed in particular cases.   The examples of Sec.~\ref{Examples} should help clarify this.   We also assume  $\Om$ denotes  an arbitrary domain of $\mathbb{R}^n$ with $\partial\Omega$ being its boundary. For convenience, we will omit the incremental volume element $d^n z$ for integrations over $\Omega$, i.e., $\int_\Om= \int_\Om d^n z$. We assume strong boundary conditions such that all integrations by parts produce vanishing boundary terms.  

Given our choice of $\xi^N$ as the entropy density, the  total entropy is evidently given by the  following:
\bq
 S[\bfxi] = \int_\Om \xi^N\,.
 \label{totS}
\eq 
This functional is required to be a Casimir invariant of the noncanonical Poisson bracket $\{\,\cdot,\cdot\,\}$, which one is assumed to have found in  the third step of the UT algorithm, i.e., 
\bq
\{F,S\}=0, \qquad \forall F\,.
\eq
The Hamiltonian functional $H$ associated to the noncanonical Poisson bracket $\{\,\cdot,\cdot\,\}$ is  given by
\bq
H[\bfxi] = \int_\Om h\,, \label{totH}
\eq
where   $h$, the Hamiltonian density,  in general depends on all the variables $\xi^1, \xi^2, \ldots, \xi^{N}$.  We will take  $H$ to be the total energy, as is  indeed the case for the examples mentioned in  Sec.~\ref{Hamiltonian formalism}. The evolution of the dynamical variables in the ideal case, i.e., when dissipation is not included, is given by
\bq
\partial_t \xi^{\al} = \{\xi^\al,H\},\quad  \al = 1,2,\ldots,N\,.
\label{evolham}
\eq
Now it remains to add to  \eqref{evolham} the dissipative evolution, which has  the following natural combined form:
\bal
\partial_t\xi^\al &= \{\xi^\al,H\} + \call^{(\al)}\!\cdot\bfJ^{\al}\,,\quad  \al=1,\ldots,N-1\,,
\label{N-1-evol}
\\
\partial_t\xi^N &=  \{\xi^N,H\} +\call^{(N)}\!\cdot\bfJ^{N} +   \bfZ_{\al}\cdot \tilde{L}^{\al\be}\cdot \bfZ_{\be}\, .
\label{N-entropy}
\eal
 Equation \eqref{N-1-evol} is the sum of two conservative terms, the first being Hamiltonian, while the second is dissipative. In this second expression  $\al$ is not summed, but a particular operator $\call^{(\al)}$  may  act  on each flux $\bfJ^\al$.  Recall, this was the purpose of the parenthesis.  If $\xi^\al$ were a rank $m$ tensor, then usually $\bfJ^\al$ would be of rank $m+1$ with the contraction indicated by ``$\, \cdot\,$" providing tensorial consistency.  However, we leave open the possibility that  $\call^{(\al)}$  may contribute to tensorial consistency.  For usual nonequilibrium thermodynamics $\call^{(\al)}=-\nabla$,  for all $\al$,  and the conservative form of \eqref{N-1-evol} is manifest. Equation \eqref{N-entropy} similarly has conservative terms, but the addition of the last term is responsible for entropy production.  Because $\xi^N $ is a scalar density, $\bfJ^N$ is a vector and the contractions of $\bfZ_{\al}\cdot \tilde{L}^{\al\be}\cdot \bfZ_{\be}$  between some ``vector fields"   $\bfZ_{\al}$  and a quantity $\tilde{L}^{\al\be}$ produces   the correct tensorial form.  Since the entropy production must be guaranteed, we assume
$\tilde{L}^{\al\be}$ is  symmetric and positive semidefinite, giving 
\bq
\dot{S}  =\int_\Om\bfZ_{\al}\cdot \tilde{L}^{\al\be}\!\cdot \bfZ_{\be} \eqqcolon \int_\Om \dot{s}^{prod}\geq 0\,.
\label{dot_s_prod}
\eq

The construction above is similar to that  presented in \citep[][see page 14]{coquinot2020general},  in order to construct a general form of metriplectic 2-bracket.  However, there the  pseudodifferential operators were all taken to be spatial gradients, i.e.,  $\call^{(\al)}\coloneqq -\nabla$.  Here we generalize this by supposing  each operator $\call^{(\al)}$ has an adjoint ${\call_{*}^{(\al)}}$ defined with respect to the standard inner product, i.e.,  $(f,g) = \int_\Om f\,g$, which of course is  the case for $\nabla$ where  $\nabla_* = -\nabla$.

What we have accomplished so far is the first step of  the {\it anlysis-synthesis} method, the  {\it analysis phase}.    With this method we work   backwards from the desired form of the dynamical equations \eqref{N-1-evol} and \eqref{N-entropy}.  In the second step, the {\it synthesis phase}, we determine explicitly the quantities $\bfJ^{\al}$, $\bfZ_\al$ and $\tilde{L}^{\al\be}$. We will show that these quantities are expressed in terms of the functional derivatives of the Hamiltonian,  $H_{\xi^\al}$.  To be clear,  we remind the reader that the goal of this analysis-synthesis process is to construct the operators  $M$ and $\Si$.
 
Given any functional $F[\bfxi]$, we have the basic identity
\bq
\dot F[\bfxi]=\int_\Om \frac{\de F}{\de \xi^\al} \, \p_t \xi^\al\,.
\label{BId}
\eq
This follows upon assuming $\Om$ is fixed and boundary terms vanish, which we have assumed throughout. 
Applying \eqref{BId} to $H$ and using our  notation  $H_{\xi^{\al}}= \de H/\de \xi^{\al}$,  we obtain upon substitution of \eqref{N-1-evol} and \eqref{N-entropy}  
\bal
\dot H[\bfxi]&= \int_\Om  H_{\xi^\al} \, \call^{(\al)}\!\cdot\bfJ^{\al}
 + {H_{\xi^N}}\,\bfZ_{\al}\cdot \tilde{L}^{\al\be}\cdot \bfZ_{\be}
 \nn\\
&=  \int_\Om  \bfJ^{\al}\cdot \call_*^{(\al)} \,  {H_{\xi^\al}} 
 + {H_{\xi^N}}\,\bfZ_{\al}\cdot \tilde{L}^{\al\be}\cdot \bfZ_{\be}\,.
\label{int_auto}
\eal
To ensure energy conservation, \eqref{int_auto} must vanish.   Simple and natural choices that achieve this are the following generalized force-flux relations:
\bal
\bfZ_\al  &=  \call_{*}^{(\al)}  H_{\xi^\al}\,,
\label{Force}\\
\bfJ^\al   &= - H_{\xi^N}  \tilde{L}^{\al\be} \call_{*}^{(\be)}  H_{\xi^\be} \,.
\label{Flux-VF}
\eal
To understand these formulas consider  the standard case where $\call^{(\al)}=-\nabla$ for all $\al$.  This gives  the force-flux relations, 
\bal
\bfZ_\al  &=  \nabla  H_{\xi^\al}\,,
\label{SForce}\\
\bfJ^\al   &= - H_{\xi^N}  \tilde{L}^{\al\be} \nabla H_{\xi^\be}= -  {L}^{\al\be} \nabla H_{\xi^\be}
\label{SFlux}\,,
\eal
where in the second equality of \eqref{SFlux} we have made comparison with \eqref{genOnsag}.  Thus
\bq
 {L}^{\al\be}   = H_{\xi^N}  \tilde{L}^{\al\be}  
\label{Scomp}
\eq
and we see that the $L^{\al\be}$ of  \eqref{genOnsag} is not the same as $\tilde{L}^{\al\be}$ of \eqref{N-entropy}. If the Hamiltonian obtains its $\si$ dependence in the standard way via an internal energy function, we will see that these quantities differ by a factor of $T$, i.e., 
\bq
 \tilde{L}^{\al\be}={L}^{\al\be}/  T \,.
\eq

Now we are in position to determine the $M$ and $\Si$ of the K-N product and hence the metriplectic 4-bracket.  
We are led to  the following  choices: 
\bal
M(dF,dG) &= F_{\xi^N}\,G_{\xi^N}\,,
\label{UTAM}
\\
\Si(dF,dG) &=   \call_*^{(\al)}(F_{\xi^\al}) \tilde{L}^{\al\be} \call_*^{(\be)}(G_{\xi^\be})
\nn\\
&=  \call_*^{(\al)}(F_{\xi^\al}) \frac{{L}^{\al\be}}{H_{\xi^N}} \call_*^{(\be)}(G_{\xi^\be})
\,.
\label{UTASi}
\eal
Here we have  chosen the simplest form for  $M$, which singles out entropy, and makes the meaning of $\Si$ perspicuous.

Constructing the 4-bracket with these choices of $M$ and $\Si$, according to
\bq
(F,K;G,N) = \int_\Om \,(\Sigma \wedge M)\left(dF, dK, dG, dN\right),
\eq
gives \eqref{N-1-evol} and \eqref{N-entropy}, in  metriplectic form, viz.
\bq
\partial_t \xi^\al = \{\,\xi^\al,H\,\} + (\xi^\al,H;S,H),\quad \forall \al=1,\ldots,N\,.
\eq
Manifestly, \eqref{dotH1} is satisfied and, we have for \eqref{dotS1}
\bal
\dot{S} &=   (S,H;S,H) =K(H,S)= \int_\Om \Si(dH,dH)\nn \\
&=\int_\Om \call_*^{(\al)} (H_{\xi^\al}\!)\, \tilde{L}^{\al\be} \call_*^{(\be)} (H_{\xi^\be}\!)\geq 0\,.
\label{Sprod1}
\eal
Comparison with \eqref{dot_s_prod} reveals $\dot{s}^{prod}$ becomes 
\bal
\dot{s}^{prod} &= \Si(dH,dH) = \call_*^{(\al)} \left( H_{\xi^\al} \right) \tilde{L}^{\al\be} \call_*^{(\be)} (H_{\xi^\be})
\nn\\
&= \call_*^{(\al)} \left( H_{\xi^\al} \right) \frac{{L}^{\al\be}}{H_{\xi^N}}\call_*^{(\be)} (H_{\xi^\be})
\,.
\label{Sprod2}
\eal
Thus,   the theory is complete once the phenomenological coefficients $L^{\al\be}$ are determined.  We reiterate  that our construction clearly delineates between the phenomenological laws embodied in  $L^{\al\be}$ and  the local thermodynamics contained in the Hamiltonian, e.g.,  in the internal energy function.  Also,  choosing $M$ as in \eqref{UTAM} endows $\Si$ with the physical meaning  inherent in \eqref{Sprod1}  and \eqref{Sprod2} relating entropy production and sectional curvature.

 We comment further on these coefficients in the context of non-equilibrium thermodynamics theory  in  Sec.~\ref{non-equilibrium thermodynamics theory}.

\subsection{Non-equilibrium thermodynamics theory}
\label{non-equilibrium thermodynamics theory}

Many phenomena can be described by the idea that  fluxes are caused by gradients of quantities, which are viewed as the thermodynamic forces.   For example,  Fourier's law relates heat flow to temperature gradients,  Fick’s law relates diffusion to concentration gradients, and in the Navier-Stokes equation  momentum flux  is related to velocity gradients. In non-equilibrium thermodynamics this is generalized by assuming   fluxes  are  linear combinations  of thermodynamic forces and thereby allowing for cross-effects.  This is the essence of  the  Onsager reciprocal relations \cite{Onsager,Casimir},  which are here represented  by the   force-flux relations of  \eqref{onsager1}   (see, e.g., \cite{de2013non}).

For gaseous systems, an underlying  kinetic theory can provide a justification for the phenomenological  relations embodied in  the $L^{\al \be}$.  This is the case for  low-density gases, but in general  such calculations are difficult or even prohibitive.   However, many irreversible processes are empirically  seen to be governed by linear relations between fluxes and  forces \cite{Miller} and in this way the  $L^{\al \be}$ are provided.  However they are provided,  our theory leaves open the possibility  that they can depend on all the dynamical variables.

Returning to the theory developed in Sec.~\ref{K-N Construction and Fluxes}, we observe from equation \eqref{Flux-VF} that the thermodynamic force-like terms now take the new form  $\call^{(\al)}  H_{\xi^\al}$, where $H$ is the Hamiltonian functional (cf.~Eq.~\eqref{genOnsag}). In Sec.~\ref{Examples}  we  will confirm that our new form $\call^{(\al)}  H_{\xi^\al}$ can match  known examples and  that our  last step  of the UT algorithm leading to the metriplectic 4-bracket provides a means for generalizing known examples and providing new thermodynamically consistent theories.

\section{Examples}
\label{Examples}

In this section, we will give three examples. For all the examples, we consider the case were we have a single real valued field variable depending on one space- and one time-independent variable, $\bfxi(\bfx,t)$ where $\bfx = (x^1,x^2,x^3)$ is a Cartesian coordinate for a  fluid contained in a  volume  $\Om$. Throughout the following, we use boldface to denote vectors, an over bar to denote rank-2  tensors, and a double over bar to denote rank-4 tensors.

\subsection{Navier-Stokes-Fourier (NSF)}
\label{NSF}

We proceed with the UT algorithm motivated by  our previous development of the NSF \cite{pjmU24,zaidni2};  we find the algorithm produces a more general system that contains the NSF as a special case.

    \bigskip

 \noindent$\bullet$   \underline{First step of UT algorithm}: We choose the set of fluid variables as follows: 
 \bq
 \bfxi(\bfx,t) = (\rho(\bfx,t),\bfm(\bfx,t),\si(\bfx,t))\,,
 \eq
where $\rho$ is the mass density, $\bfm=\rho\bfv$ is the momentum density with $\bfv$ being the Eulerian velocity field, and $\si$ is the entropy density. Observe we have singled out the entropy density $\si$ as the last variable, consistent with  \eqref{xi_x}. 
 
 \medskip
 
 \noindent$\bullet$   \underline{Second step of UT algorithm}: Consistent with  \eqref{totS}, we take the total entropy to be the integral of the last component
    \bq 
    S = \int_\Om \si
    \label{S_nsf}\,.
    \eq
 A natural choice of  Hamiltonian functional for NSF  is 
      \bq
     H = \int_\Om \frac{\, |\mathbf{m}|^2}{2\rho} + \rho\, u(\rho,\si/\rho)\,,  
     \label{Ham_1}
     \eq
the sum of fluid kinetic energy and $\rho$ times the specific internal energy $u$, which is known to be conserved by the NSF.  More general Hamiltonians including, e.g.,  the  gravitational force would be straightforward.
 The usual thermodynamic  relations are
\bq
p=\rho^2\frac{\p u}{\p \rho} \qquad \mathrm{and}\qquad T=\frac{\p u}{\p s}\,,
\label{thermo1}
\eq
where the specific entropy $s=\si/\rh$.  Alternatively, we can leave the Hamiltonian unspecified, i.e., let it be any functional $H[\rho,\bfm,\si]$ -- independent of its form any   $H$ will be conserved by the metriplectic 4-bracket dynamics. 
    
    \medskip
    
 \noindent$\bullet$    \underline{Third step of UT algorithm}: The appropriate Poisson bracket is the so-called Lie-Poisson bracket  given in \cite{pjmG80}.  For  two functionals $F,G\in \calb$ it is defined as follows:   
    \bal
    \{F,G\} = -  \int_{\Omega} \mathbf{m}&\cdot \left[  F_{\bfm} \cdot \nabla G_{\bfm} - G_{\bfm}\cdot \nabla F_{\bfm}\right]
     \nn\\
     &+ \rho\left[F_{\bfm}\cdot \nabla G_\rho  -G_{\bfm}\cdot \nabla F_\rho  \right]
     \nn\\
     &+ \sigma \left[F_{\bfm}\cdot \nabla G_\si - G_{\bfm}\cdot\nabla  F_\si\right] \,,
     \label{PB_NSF}
     \eal
     where $S$ is a Casimir invariant, i.e., $\{\,S,F\,\} = 0$, for any functional $F$. 
     
     \medskip
     
 \noindent$\bullet$     \underline{Fourth step of UT algorithm}: To construct the metriplectic 4-bracket, we proceed with the systematic development  presented in Sec.~\ref{K-N Construction and Fluxes}, viz., $M$ and $\Si$ are given by
 \bal
M(dF,dG) &= F_{\si}\,G_{\si}\,,
\label{UTANSFM}
\\
\Si(dF,dG) &=  \call_*^{(\al)}(F_{\xi^\al}) \frac{{L}^{\al\be}}{H_{\si}} \call_*^{(\be)}(G_{\xi^\be})
\,,
\label{UTANSFSi}
\eal
and the UT algorithm is complete up to the choices for $\call_*^{(\al)}$ and ${L}^{\al\be}$.  For any choices of these quantities, according to \eqref{Flux-VF} and \eqref{Scomp}, the 4-bracket using \eqref{UTANSFM} and \eqref{UTANSFSi} will be consistent with the following general expressions for the fluxes:
\begin{small}
 \bal
     {\bfJ}_\rho &=-  L^{\rho\rho}\!\cdot \call_*^{\rho}(H_\rho) - L^{\rho\bfm}\!:\!\call_*^{\bfm}(H_\bfm) 
   - L^{\rho\si}\!\cdot \call_*^{\si}(H_\si)\,, 
    \nn \\
     \bar{J}_{\mathbf{m}} &=-  L^{\bfm\rho}\otimes\call_*^{\rho}(H_\rho) - L^{\bfm\bfm}\!:\!\call_*^{\bfm}(H_\bfm) - L^{\bfm\si}\otimes\call_*^{\si}(H_\si) \,,\nn
     \\
     {\bfJ}_s &=-   L^{\si\rho}\!\cdot\call_*^{\rho}(H_\rho) -  L^{\si\bfm}\!:\!\call_*^{\bfm}(H_\bfm) - L^{\si\si}\!\cdot \call_*^{\si}(H_\si) \,, 
     \label{genNSFflux}
     \eal
 \end{small}where  ${\bfJ}_\rho$ is the net mass flux, ${\tensJ}_{\mathbf{m}}$ is the momentum flux, and ${\bfJ}_s$ is the net entropy flux. Thus we have obtained a quite general  thermodynamically consistent system, one that generalizes  the NSF system.  In fact, the 4-bracket that produces \eqref{genNSFflux} is sufficiently general to produce  the Brenner-Navier-Stokes system of Sec.~\ref{BNSF} and the significant  generalizations of the BNS that we describe there.

Now we specialize and show that the general expressions for the fluxes of \eqref{genNSFflux} reduce to those known for the NSF (see, e.g.,   \cite{de2013non,pjmC20,pjm84b}), viz.
\bq
    {\bfJ}_\rho = 0\,,
\quad
     \bar{J}_{\mathbf{m}} =  - \4tensLa:\nabla \mathbf{v}\,,
\quad
     {\bfJ}_s = - \frac{\tenska}{T}\cdot\nabla T\,,
     \label{NSFflux}
\eq
where  ${\bfJ}_\rho$ is the net (vector) mass flux, ${\tensJ}_{\mathbf{m}}$ is the momentum flux (rank 2 tensor), and ${\bfJ}_s$ is the net (vector) entropy flux. In \eqref{NSFflux} $\tenska$ is the  thermal conductivity tensor, $\tensD$ is the  diffusion tensor, which along with $\tenska$ is assumed to be a symmetric and positive definite  2-tensor,  and 
$\4tensLa$ is the viscosity 4-tenor, the usual rank 4 isotropic Cartesian tensor  given by
\bq
\La_{ijkl}=\et\left(\de_{il} \de_{jk}+ \de_{jl}\de_{ik} -\frac2{3} \de_{ij}\de_{kl} \right) + \ze\, \de_{ij}\de_{kl}\,,
\label{viscous}
\eq
with viscosity coefficients $\et$ and $\ze$ and $i,j,k$ and $l$ taking on values 1,2,3.   In \eqref{NSFflux} and henceforth we use a  single ``$\,\cdot\,$" to indicate neighboring contractions and we use the double dot convention as follows: 
\bal
(\tenska\cdot \nabla G_{\si})_i&=\ka_{ij} \p_j G_{\si}\nn\\
(\4tensLa:\!\nabla \bfm)_{ij}&=\La_{ijkl} \p_k m_l\nn\\
(\bm{\ep}:\!\nabla \bfm)_{i}&=\ep_{ijk} \p_j m_k
\label{bep}
\eal
 where repeated indices are summed over.  We have added  \eqref{bep} for later use, when we have a  double  contraction with  a   3-tensor $\bm{\ep}$.

To see how the fluxes of \eqref{NSFflux} emerge from our general expressions of  \eqref{genNSFflux}  we  set  $\call_*^{(\al)}=\nabla$, for all $\al$,  and assume $H$ is given by \eqref{Ham_1};  therefore 
    \bq 
    H_\si = T,\quad H_\bfm = \bfv,\quad H_{\rho} = -\frac{|\bfm|^2}{2\rho^2} - \frac{T\,\si}{\rho} + \frac{p}{\rho} + u,
    \label{HdivNSF}
    \eq
 and comparison of  \eqref{genNSFflux} with \eqref{NSFflux} reveals that the only nonzero phenomenological coefficients $L^{\al\be}$ are the following:
 \bq
  L^{\bfm\bfm} =  {\4tensLa}  \andq L^{\si\si} = \frac{\tenska}{T}\,.
\eq
     
Thus we immediately obtain  $\Si$  from \eqref{UTANSFSi} as 
\bal
     {\Si}(dF ,dG ) &=  \nabla F_{{\bfm}} : \frac{L^{\bfm\bfm}}{H_\si} : \nabla G_{{\bfm}} + \nabla F_{{\si}} \cdot \frac{L^{\si\si}}{H_\si} \cdot  \nabla G_{{\si}} 
     \nn\\
      &=  \nabla F_{{\bfm}} : \frac{\4tensLa}{T} : \nabla G_{{\bfm}} + \nabla F_{{\si}} \cdot \frac{\tenska}{T^2} \cdot  \nabla G_{{\si}} 
     \label{SiSI_nsf}
\eal
which together with the expression for $M$ of \eqref{UTANSFM} gives 
 the 4-bracket  
     \bal
     &(F, K ; G, N)= 
      \label{4BSINSF}
       \\
      &\hspace{.3cm} \int_{\Omega}\!\frac{1}{T}\Big[
     \left[K_\sigma \nabla   F_{\mathbf{m}}-F_\sigma \nabla   K_{\mathbf{m}}\right]
     \!:\!\4tensLa \!:\!
     \left[N_\sigma \nabla   G_{\mathbf{m}}-G_\sigma \nabla   N_{\mathbf{m}}\right]
     \nn\\
    &\hspace{.5cm} +\ \frac{1}{T}\,
     \big[K_\sigma \nabla  F_{{\sigma}}-F_\sigma \nabla  K_{{\sigma}}\big]
     \!\cdot \!\tenska\! \cdot \!\big[N_\sigma \nabla  G_{{\sigma}}-G_\sigma \nabla  N_{{\sigma}}\big]  \Big]
     \,.
     \nonumber
     \eal
Insertion of  $H$ of  \eqref{Ham_1}  and $S$ of \eqref{S_nsf}, yields the  NSF dynamical system
    \bal
   \p_t \rho  &=\{\rho,H\}+ (\rho,H;S,H)
    \nn\\
    &= -\bfv\cdot\nabla\rho   - \rho\, 
     \nabla\cdot\mathbf{v} 
      \,,
      \label{DprtNSF}
      \\
       \p_t \mathbf{v}  &=\{\bfv,H\}+ (\bfv,H;S,H)
    \nn\\
    &= -\bfv\cdot\nabla \bfv -\nabla p/\rho
   + \frac1{\rho}\nabla\cdot(\4tensLa:\nabla \mathbf{v})\,,
   \label{DpvtNSF}
   \\
     \p_t \si &=\{\si ,H\}+ (\si,H;S,H)
     \nn\\
     &= - 
      \bfv\cdot\nabla\si 
     - \sigma\,\nabla\cdot\mathbf{v}
     + \nabla\cdot\left(\frac{\tenska}{T}\cdot\nabla T\right)
       \nn\\ 
       &\hspace{1.25cm} + \frac{1}{T^2}\nabla T\cdot\tenska \cdot \nabla T 
+ \frac{1}{T}\nabla\mathbf{v}:\4tensLa :  \nabla\mathbf{v}\,.
      \label{Dpst}
      \eal 
       By construction we automatically have the entropy production 
      \bq
      \dot{{S}}= ({S}, {H} ; {S}, {H})
       = \int_\Om \dot{s}^{prod} \geq 0\,.
     \label{dots2}
     \eq
     where 
     $$\dot{s}^{prod}= \nabla\mathbf{v}: \frac{\4tensLa}{T} :\nabla\mathbf{v} 
     + \nabla T \cdot \frac{\tenska}{T^2}\cdot \nabla T\, .
     $$
 It is important to note that the square of the temperature in the denominator of the coefficient $\tenska/T^2$ in $\dot{s}^{prod}$, has factors from different physical origins. One factor  comes from the systematic theory, where temperature is defined as $T\coloneqq H_\si$, while the second arises from the phenomenological law, specifically Fourier's law, where the heat flux is given by  $\bfq = -\tenska \, \nabla T/T$.

\subsection{Cahn-Hilliard-Navier-Stokes (CHNS)}
\label{CHNS}

 Various equations have been proposed for describing two-phase fluid flow by combing the physics of the Cahn-Hillard equation \cite{cahn1958free} with that of the Navier-Stokes equations (see, e.g., \cite{anderson2000phase,anderson98,anderson96,Guo} and references therein).   In these CHNS models the influence of a second phase of matter is included by adding a concentration variable that describes the second phase.  In \cite{zaidni2} we used the metriplectic 4-bracket formalism to obtain a general model that encompasses, corrects, and generalizes existing models.   As in Sec.~\ref{NSF}, in this section we will proceed with the UT algorithm and obtain a very general thermodynamically consistent two-phase flow system.

    \bigskip
 
  \noindent$\bullet$     \underline{First step of UT algorithm}:  To our set of fluid variables we add a   variable $\tilde{c}$ which is a concentration per unit volume that describes the second phase.  Thus our dynamical variables are  
 \bq
  \bfxi(\bfx,t) = (\rho(\bfx,t),\bfm(\bfx,t),\tilde{c}(\bfx,t),\bar{\si}(\bfx,t))\,,
 \eq
where again the mixture of two phases is assumed to be contained in a volume  $\Om$, with coordinate $\bfx$,  and to  the densities $\rho,\bfm$, and $\bar\si$  used as  in Sec.~\ref{NSF} we add $\tilde{c}$.    Again we have singled out the entropy density $\bar\si$ as the last variable of $\bfxi$, consistent with  \eqref{xi_x}.  (Note, the  reason for the bar  will soon become clear.)  The specific  concentration associated with $\tilde{c}$ is given by ${c}=\tilde{c}/\rho$.

  \medskip
 
 \noindent$\bullet$    \underline{Second step of UT algorithm}:   Again, consistent with  \eqref{totS}, we take the total entropy to be the integral of the last component
    \bq 
    S = \int_\Om \bar\si
\label{S_CHNS}\,.
    \eq
 It was shown in \cite{zaidni2} that this simple entropy can be used instead of complicated  entropy expressions  used in \cite{anderson2000phase,anderson98,anderson96,Guo}, which were modeled after the free energy of the Cahn-Hilliard equation. 
    
We record here for later use  the relationship between our simple entropy $\bar{\si}$ and the previous one which we denote  by $\bar{\si}$, viz. 
\bq
\bar{\si} =  {\si} + \frac{\la_s}{2}\, {\Ga^2(\na {c})}\,.
\label{entMap}
 \eq
Here  the coefficient $\lambda_s$ is a constant and the function $\Gamma$ is a homogeneous function of degree unity, i.e., 
\bq
\Gamma (\lambda \bfp ) = \lambda \Gamma(\bfp ) \text{  for all   } \lambda > 0\,.
\label{Ehomo1}
\eq
 Because $\Gamma$ is a homogeneous function of degree unity we have
 \bq
 \Gamma(\bfp ) = \bfp\cdot \bfze \coloneqq p_j  \frac{\partial \Gamma(\bfp)}{\partial p_j}\,.
 \label{GF1}
 \eq
where $\bfze$ is a  homogeneous function of degree zero.  The function $\Ga$ was shown in  \cite{taylor92} to describe anisotropic weighted mean curvature effects due to anisotropic surface tension. 

Any Hamiltonian $H[\rho,\bfm,\tilde{c},\bar{\si}]$ would be possible;  however, as also shown in \cite{zaidni2},  the price paid for a simplified entropy is the following complicated Hamiltonian:
\bq
     H = \int_\Om \!\frac{\, |\mathbf{m}|^2}{2\rho} 
     + \rho\,  u\big(\rho, {\si}/\rho,\tilde{c}/\rho\big) 
+ \frac{\la_u}{2}\, \Ga^2\big(\na (\Tilde{c}/\rho)\big)\,,
     \label{Ham_CHNSA}
\eq
where in the second argument of the internal energy $u$ we have inserted $\si$ as a shorthand for the expression in terms of $\bar{\si}$, $\tilde{c}$ and $\rho$ obtained upon inserting  $\si$ from \eqref{entMap}. From this extensive internal energy function, we obtain  the intensive thermodynamical variables including the chemical potential   as 
\bq
p=\rho^2\frac{\p u}{\p \rho}\,, \qquad  T=\frac{\p u}{\p s}\,, \andqq
\mu=\frac{\p u}{\p c}\,.
\label{TH2}
\eq
where now $s=\bar{\si}/\rho$ and recall $c=\tilde{c}/\rho$.  The parameter  $\lambda_u$ is another constant that describes anisotropic surface energy effects.  In the previous works it was shown how the constants $\la_s$ and $\la_u$ are related to   $\lambda_f$,  a parameter that depends  on the temperature according to  
\bq
\la_f(T) = \la_u - T \la_s\,.
\eq
We take this as given, referring the reader to the previous works for explanation. 
 
   \medskip
 
 \noindent$\bullet$    \underline{Third step of UT algorithm}: The appropriate Poisson bracket, defined on two functionals $F,G\in \calb$,  is that for the Gibbs-Euler system given in \cite{zaidni2}.  This bracket, which is a  natural generalization of that given in \cite{pjmG80}, is given by   
\bal
\{F,G\} = -  \int_{\Omega}& \mathbf{m}\cdot \left[  F_{\bfm} \cdot \nabla G_{\bfm} - G_{\bfm}\cdot \nabla F_{\bfm}\right]
 \nn\\
&+ \rho\left[F_{\bfm}\cdot \nabla G_\rho  -G_{\bfm}\cdot \nabla F_\rho  \right]
 \nn\\
&+ \bar{\si} \left[F_{\bfm}\cdot \nabla G_{\bar{\si}} - G_{\bfm}\cdot\nabla  F_{\bar{\si}}\right] 
 \nn\\
&+ \Tilde{c} \left[F_{\bfm}\cdot \nabla  G_{\Tilde{c}} - G_{\bfm}\cdot\nabla    F_{\Tilde{c}} \right] \,.
\label{PB_CHNS}
\eal
It is simple to verify that the $S$ of \eqref{S_CHNS} is a Casimir invariant of this bracket. 

     \medskip
     
 \noindent$\bullet$     \underline{Fourth step of UT algorithm}: To construct the metriplectic 4-bracket, we proceed as in Sec.~\ref{NSF}
 with the forms of $M$ and $\Si$ given by \eqref{UTANSFM} and \eqref{UTANSFSi}, albeit with $\bar{\si}$ replacing $\si$ in  \eqref{UTANSFM}.  Thus the determination of our system is  complete  when we make choices for $\call_*^{(\al)}$ and the  ${L}^{\al\be}$.   For any choices of these quantities,  the 4-bracket constructed from $M$ and $\Si$  will be consistent with the following general expressions for the fluxes obtained from \eqref{genOnsag}:
\begin{small}
 \bal
     {\bfJ}_\rho &=- L^{\rho\rho}\!\cdot \call_*^{\rho}(H_\rho)  - L^{\rho\bfm}\!: \call_*^{\bfm}(H_\bfm)  \nn\\
     &\hspace{2.25cm}-  L^{\rho\si}\!\cdot \call_*^{{\bar{\si}}}(H_{\bar{\si}})-  L^{\rho \Tilde{c}}\!\cdot \call_*^{\tilde{c}}(H_{\tilde{c}})  \,,\nn
     \\
     \bar{J}_{\mathbf{m}} &=-  L^{\bfm\rho}\otimes\call_*^{\rho}(H_\rho) - L^{\bfm\bfm}:\call_*^{\bfm}(H_\bfm)  \nn\\
     &\hspace{2.25cm}-  L^{\bfm\si}\otimes \call_*^{{\bar{\si}}}(H_{\bar{\si}})  - L^{\bfm \Tilde{c}}\otimes\call_*^{\tilde{c}}(H_{\tilde{c}}) \,,\nn
     \\
     {\bfJ}_c &=-   L^{\tilde{c}\rho}\!\cdot \call_*^{\rho}(H_\rho) - L^{\tilde{c}\bfm}\!:\call_*^{\bfm}(H_\bfm) \nn\\
     &\hspace{2.25cm} - L^{\tilde{c}\bar{\si}}\!\cdot \call_*^{{\bar{\si}}}(H_{\bar{\si}}) - 
     L^{\Tilde{c} \Tilde{c}}\!\cdot \call_*^{\tilde{c}}(H_{\tilde{c}}) \,,\nn
     \\
     {\bfJ}_s &=-   L^{{\bar{\si}}\rho}\!\cdot \call_*^{\rho}(H_\rho) - L^{{\bar{\si}}\bfm}\!:\call_*^{\bfm}(H_\bfm)  \nn\\
     &\hspace{2.25cm} - L^{{\bar{\si}}{\bar{\si}}}\!\cdot \call_*^{{\bar{\si}}}(H_{\bar{\si}}) - L^{{\bar{\si}} \Tilde{c}}\!\cdot \call_*^{\Tilde{c}}(H_{\tilde{c}})\,. 
     \label{genCHNSflux}
     \eal
\end{small}Thus, we have obtained a quite general  class of thermodynamically consistent systems, one that generalizes  a variety of  existing CHNS systems depending on the choice of $H$,  $L^{\al\be}$,  and $\call_*^{\al}$.

 Now we specialize and show that the general expressions for the fluxes of \eqref{genCHNSflux} reduce to those known for the CHNS.  For example,  if we choose $\call_*^{(\al)}=\nabla$, for all $\al$, and $H$ to be the expression of   \eqref{Ham_CHNSA},  then we obtain the CHNS system of  Anderson et al.\ \cite{anderson2000phase,anderson98,anderson96} (see also \cite{zaidni2}).  Using 
 \bal
H_\rho &= -\frac{|\bfm|^2}{2 \rho^2}  +u+\rho u_\rho-\left(\frac{\sigma}{\rho}-\frac{\lambda_s}{2\rho}   \Gamma^2\right) u_s-\frac{\tilde{c}}{\rho} u_c \nn \\
&\hspace{1.5cm}  +\frac{\tilde{c}}{\rho^2} \nabla \cdot\left(u_s  \lambda_s \Ga\bfze\right)+\frac{\tilde{c}}{\rho^2} \nabla \cdot\left( \Ga \bfze \lambda_u\right)
\label{H_rho}
\\
H_{\tilde{c}}&=u_c+\frac{\lambda_s}{\rho} \nabla \cdot\left( u_s \Ga \bfze\right)-\frac{1}{\rho} \nabla \cdot\left(\lambda_s \Ga \bfze \right)\eqqcolon \mu_\Ga\,,
\label{H_c}
\\
H_\bfm &= \bfv, \qquad H_{\bar{\si}} = u_s = T\,.
\label{H_m_si}\,,
\eal
where, from \eqref{TH2}, we defined $u_\rho :=  {\p u}/{\p \rho}=  {p}/{\rho}$, $ u_s :=  {\p u}/{\p s} = T$ and $u_c := {\p u}/{\p c} = \mu$.  Upon setting all the $L^{\al\be}$ to zero except 
 \bq
     L^{\bfm\bfm} =  {\4tensLa} \,,\quad   L^{\bar{\si}\bar{\si}} =   \frac{\tenska}{T}\,, 
\andq L^{\Tilde{c}\Tilde{c}}  =  {\tensD}\,.
     \eq
Equations \eqref{genCHNSflux} for the fluxes reduce to the  following form:
\bal
    {\bfJ}_\rho &= 0\,,\label{mass_flux_CHNS}
\qquad 
     \bar{J}_{\mathbf{m}} =  - \4tensLa:\nabla \mathbf{v}\,,
\\
     {\bfJ}_c &= - {\tensD}\cdot\nabla \mu_\Ga\,,
\qquad 
     {\bfJ}_s = - \frac{\tenska}{T}\cdot\nabla T\,,\label{entropy_flux_CHNS}
\eal
where $\mu_\Ga := \mu - \frac{1}{\rho}\nabla\cdot(\la_f \Ga \bfze)$ and $\bar{D}$ is a rank-2 diffusion tensor. Equations are the known fluxes for the CHNS system of \cite{anderson2000phase,anderson98,anderson96}.

The metriplectic 4-bracket for this case, as determined by 
     \bal
      M(dF ,dG ) &= F_{\bar{\si}}G_{\bar{\si}}\,,
     \\
      {\Sigma}(dF ,dG ) &=  \nabla F_{{\bfm}} : \4tensLa: \nabla G_{{\bfm}} +  \nabla F_{\bar{\si}} \cdot \frac{\tenska}{T^2}\cdot \nabla G_{\bar{\si}} 
       \nn\\
     &\hspace{1.5cm} + \na(F_{\Tilde{c}}) \cdot\frac{\tensD}{T}\cdot \na(G_{\Tilde{c}}) \,,
     \label{SiSI3CHNS}
     \eal
is 
     \bal
     (F, K ; G, N)&= 
     \label{4BDICHNS}\\
     & \hspace{-1.9cm} \int_{\Omega}\!\frac{1}{T}\Big[
     \left[K_{\bar{\si}} \nabla   F_{\mathbf{m}}-F_{\bar{\si}} \nabla   K_{\mathbf{m}}\right] \!
     \colon\!\! \4tensLa \colon\!\!
     \left[N_{\bar{\si}} \nabla   G_{\mathbf{m}}-G_{\bar{\si}} \nabla   N_{\mathbf{m}}\right]
     \nn\\
     &\hspace{-1.88cm}+ \frac{1}{T}
     \big[K_{\bar{\si}} \nabla  F_{{\bar{\si}}}-F_{\bar{\si}} \nabla  K_{{\bar{\si}}}\big]
     \cdot\tenska\cdot \big[N_{\bar{\si}} \nabla  G_{{\bar{\si}}}-G_{\bar{\si}} \nabla  N_{{\bar{\si}}}\big]
     \nn\\
     &\hspace{-1.88cm}+ 
     \!\big[K_{\bar{\si}} \na F_{\tilde{c}}  -F_{\bar{\si}}  \na K_{\tilde{c}}\big]\!\cdot\! \tensD\!\cdot\!
     \big[N_{\bar{\si}} \na G_{\tilde{c}} - G_{\bar{\si}}\na N_{\tilde{c}}\big].
     \nn
     \eal
     Upon insertion of  $H$  given by \eqref{Ham_CHNSA}  and $S$ given by \eqref{S_CHNS}, using 
 \eqref{H_rho}, \eqref{H_c}, and \eqref{H_m_si} with $S_{\bar{\si}} = 1$, the following CHNS system is produced
    \bal
    \p_t \rho  &=\{\rho,H\}+ (\rho,H;S,H)
    \nn\\
    &= -\bfv\cdot\nabla\rho   -\rho\, \nabla\cdot\mathbf{v} \,, 
      \label{Dprt30}
      \\
  \p_t \mathbf{v} &=\{\bfv,H\}+ (\bfv,H;S,H)
   \nn\\
   &= -\bfv\cdot\nabla \bfv -\frac{1}{\rho}\nabla \cdot \Big[\left(p -  \lambda_f\Gamma^2/2\right)\bar{I}
    \nn\\
    & \hspace{1cm}+\lambda_f \Gamma{\bfze}\otimes\nabla c\Big]
 + \frac1{\rho}\nabla\cdot(\4tensLa:\nabla \mathbf{v})\,,
   \label{Dpvt30}
\\
    \p_t \Tilde{c}  &=\{\tilde{c},H\} + (\tilde{c},H;S,H)
     \nn\\
    &= -\bfv\cdot\nabla\tilde{c} -\Tilde{c}\,\nabla\cdot\mathbf{v}
 + \nabla\cdot(\tensD\cdot\nabla\mu^0_\Ga)\,,
      \label{Dpct30}
    \\
    \p_t \bar{\si}  &=\{\bar{\si}  ,H\} + (\bar{\si}  ,H;S,H)
      \nn\\
    &= -\bfv\cdot\nabla \bar{\si}  
     -  \bar{\si}  \,\nabla\cdot\mathbf{v}
           \label{Dpst30}
           \\ &\quad + \nabla\cdot\left(\frac{\tenska}{T}\cdot\nabla T\right) + \frac{1}{T^2}\nabla T\cdot \tenska\cdot\nabla T  
            \nn\\
    &\hspace{1cm}+ \frac{1}{T}\nabla\mathbf{v}:\4tensLa : \nabla\mathbf{v} + \frac{1}{T} 
       \nabla\mu_\Ga\cdot \tensD \cdot \nabla\mu_\Ga \,, 
\eal 
where $\bar{I}$ is the identity and recall $\bfxi$ is defined in \eqref{GF1}, 
\bq
 \mu_\Ga\coloneqq u_c+\frac{\lambda_s}{\rho} \nabla \cdot\left( u_s \Ga \bfze\right)-\frac{1}{\rho} \nabla \cdot\left(\lambda_s \Ga \bfze \right) \,,
\eq
and $\otimes$ is the usual tensor product  $(\bfw \otimes \bfv)_{ij}= w_i v_j$.
 
 The total entropy is governed by the following: 
\bal
\dot{S}&= (S, H ; S, H) 
\nn\\
&= \int_\Om \frac{1}{T} \bigg[
\nabla\mathbf{v}:\4tensLa:\nabla\mathbf{v} + \frac{1}{T}\nabla T \cdot \tenska\cdot \nabla T 
\nn\\
&\hspace{3.10cm}+   \nabla\mu_\Ga\cdot \tensD\cdot\nabla\mu_\Ga
\bigg] \geq 0\,,
\label{dotsCHNS}
\eal
whence it is seen to be produced.

As noted above,  some previous  approaches to modeling CHNS systems employed  nonstandard    entropy expressions   \cite{anderson2000phase,anderson98,anderson96,Guo}.  In \cite{zaidni2} we proposed the following general expression, written in terms of the variables $(\rho,\bfv,c,s)$, which  encompasses  the  previous  nonstandard  expressions as special cases:
\bal
{S^a} &= \int_{\Omega} \rho s + \frac{\rho^{a} }{2} \lambda_{s} \Gamma^2(\nabla c) 
\label{CS2}
\,,\\
H^a &= \int_{\Omega} \frac{\rho}{2} |\mathbf{v}|^2 + \rho u(\rho,s,c)  + \frac{\rho^{a} }{2} \lambda_{u} \Gamma^2(\nabla c) 
\,.
\label{TE1}
\eal
Upon setting $a=0$, \eqref{CS2} and \eqref{TE1} reduce to the expressions of  \cite{anderson2000phase,anderson98,anderson96}, while upon setting $a=1$ they reduce to those of \cite{Guo} provided  the choice of an isotropic surface energy is assumed, viz.,   $\Gamma (\nabla c) = |\nabla c|$.  These were apparently  modeled after the free energy expression of the Cahn-Hilliard equation which is a linear combination of energy and entropy.

In  \cite{zaidni2}  the entropy of \eqref{CS2} was simplified to the standard form of \eqref{S_CHNS}  by a coordinate change.  This   resulted in the  more complicated internal energy function of  \eqref{Ham_CHNSA}, as compared with  \eqref{TE1},  where  in the former  the $\si$ in the argument of $u$ is  replaced by  $ {\si} =\bar{\si} -\frac{\la_s}{2}  {\Ga^2(\na {c})}$.  Given that an incremental volume of fluid contains both phases, it is perhaps not surprising that the internal energy should reflect this.

In  \cite{zaidni2}  we proceeded with the UT algorithm and obtained a metriplectic 4-bracket for a generalized  system that includes both the $a=0$ and $a=1$ cases (with a small correction to \cite{Guo}). However,  because the development of Sec.~\ref{K-N Construction and Fluxes}  was not yet available, step iv) of the algorithm required some investigation  on how to appropriately place the following pseudodifferential operator in $\Si$: 
\bq
\mathcal{L}^{\Tilde{c}}(F_{\bfxi}) := \nabla \big( F_{\Tilde{c}}   + \nabla\cdot\left(\lambda_s \Gamma{\bfze} F_{\si} \right)/\rho\big)\,. 
\label{callc}
\eq
Given that the operators $\call^{(\al)}$ can be placed at will  in the expression of $\Si$ of \eqref{UTANSFSi},   it is clear that our  new development can reproduce and generalize our previous work and produce an even more general class of thermodynamically consistent models that describe two-phase flows.  Instead of pursuing this, we will show in the next section  how the development of Sec.~\ref{NSF} produces and generalizes models by Brenner and others.

\subsection{Brenner-Navier-Stokes-Fourier (BNSF)}
\label{BNSF}

In a  series of papers \cite{BRENNER2005*,BRENNER2005,BRENNER2006,BRENNER2012} Brenner proposed a modification to address what he believed to be certain limitations of the traditional Navier-Stokes-Fourier system.  In this section we will show that his theory emerges as a special case of  our development of Sec.~\ref{NSF}.   Moreover, our theory shows the following:  how to unambiguously delineate  the dissipative dynamics from the nondisipative (Hamiltonian) dynamics;   that  generalizations of Brenner's theory by other authors are again  special cases of our theory, in particular they all emerge from  \eqref{genNSFflux};  all these theories amount to modifications of the form of dissipation in the Navier-Stokes equations.

Brenner's proposed  modification is based on a ``bivelocity theory"  that  introduces the idea of two distinct velocities: the mass velocity  $\bfv_m$, which corresponds to the conventional understanding, and a  volume velocity denoted by 
$\bfv$. In studies of classical continuum fluid mechanics, these velocities are assumed to be identical. However, Brenner argued that, in general, $\bfv_m\neq \bfv$. This hypothesis   leads to a nontraditional extension of the NSF system, known as the Brenner-Navier-Stokes-Fourier (BNSF) system, which is formulated as follows:
\bal
&\p \rho + \nabla\cdot (\rho \bfv_m) = 0\,,\\
&\p (\rho\,\bfv) + \nabla\cdot(\rho\,\bfv_m\bfv) = \nabla\cdot( - p\bar{I} + \4tensLa : \nabla \bfv)\,,\\
&\p \si + \nabla\cdot(\si \bfv_m) = \nabla\cdot\Big[\frac{\tenska}{T}\cdot\nabla T
\\  &\hspace{3.75cm}- \frac{\bfw}{T}\,   ({p+u-\rho\al}) \Big] + \dot{s}^{prod}\,.
\nn
\eal
Here, as before,  $u(\rho,s)$ is the internal energy per unit mass, $\al$ is a new unconstrained phenomenological parameter,  and $\bfw$ represent the velocity difference vector, 
$$
\bfw  = \bfv - \bfv_m.
$$
(Note, in the works of Brenner  the symbol $\bfJ$ is used for this velocity difference.)

It remains to close this system by determining $\bfw $ in terms of the dynamical variables.  In \cite{BRENNER2005*} Brenner first proposed  $\bfw = \alpha \nabla \ln(\rho)$.  Later in \cite{BRENNER2006}  and  \cite{ottinger2005,BEDEAUX2006}, using   \"Ottinger’s version of GENERIC, it was settled on the following form for $\bfw$:
\bq
\bfw=  \tilde{D}  \left(\nabla {p} - \ga\nabla {T} \right)\,,
\label{J_gen}
\eq
where  for simplicity we introduced the diffusion-like coefficient $\tilde{D} \coloneqq {D'}/{(\rho^2T})$ and  the parameter $\ga$ is  defined by  
$$
\rho\al - u = p - \ga T\,. 
$$
Thus the system contain one parameter, either $\al$ or $\ga$.  By taking $\ga = \left(\frac{\p p}{\p T}\right)_\rho$, Brenner established that the difference velocity $\bfw$ becomes 
\bq
\bfw= \frac{\tilde{D}}{\ka_T}\na \ln{\rho}\,,
\label{BrkaT}
\eq
where $\ka_T = \frac{1}{\rho}\left(\frac{\p \rho}{\p p}\right)_T$ is the coefficient of isothermal compressibility, assumed to be nonnegative. 
 In these works it is claimed that this is the most general possible constitutive equation for the velocity difference $\bfw$. However, a generalization was given in \cite{Reddy19}, which we will further generalize below using the UT algorithm.

To view  the above BNSF system in a form  adapted to the UT algorithm, we interpret  $\bfv$ to be   the usual velocity field, and  write the system in term of NSF variables $\bfxi=(\rho,\bfm = \rho\bfv,\sigma)$ as follows:
\bal
&\p_t \rho + \nabla\cdot (\rho \bfv) = \nabla\cdot(\rho\bfw)\,,
\label{rho_BNSF}\\
&\p_t \bfm + \nabla\!\cdot(\bfm\otimes \bfv) = \nabla\!\cdot\!( - p\bar{I} + \4tensLa\! : \!\nabla \bfv + \bfm\otimes\bfw)\,,\label{m_BNSF}\\
&\p_t \si + \nabla\cdot(\si \bfv) = \nabla\!\cdot\!\left[\frac{\tenska}{T}\nabla T + (\si-\ga)\bfw  \right] + \dot{s}^{prod}\,.\label{sigma_BNSF}
\eal
 Except for $\ga$ and $\bfw$ the quantities above  are defined  as for the  NSF system.  
Evidently, from \eqref{rho_BNSF}, \eqref{m_BNSF}, and \eqref{sigma_BNSF} it is seen that the  fluxes are given by the following:  
\bal
    {\bfJ}_\rho &= -\rho\bfw\,,
    \label{mass_flux_BNSF}
\\
     \bar{J}_{\mathbf{m}} &=  - \4tensLa\!:\!\nabla \mathbf{v} - \bfm\otimes\bfw\,,
     \label{mom_flux_BNSF}
\\
     {\bfJ}_s &= - \frac{\tenska}{T}\cdot\nabla T - (\si- \ga)\bfw\,, 
     \label{entropy_flux_BNSF}
\eal
which determine  the phenomenological coefficients in terms of  $\bfw$.

Given the above and the results of  Sec.~\ref{NSF}, there is no need to run through the steps of the UT algorithm:  the variables $\bfxi$ are the same, the forms of $S$ and $H$ of \eqref{S_nsf} and \eqref{Ham_1} are the same, the Poisson bracket is again the  Morrison-Greene Poisson bracket of \eqref{PB_NSF} and, the form of the operators $\call^{(\al)}$ are the same.  Thus, it only remains to determine the phenomenological coefficients and these are provided by matching  \eqref{mass_flux_BNSF},  \eqref{mom_flux_BNSF} and \eqref{entropy_flux_BNSF} with \eqref{genNSFflux}.

Comparison of  \eqref{mass_flux_BNSF} with the first equation of \eqref{genNSFflux} leads to  the determination of $\bfw$.   We have 
\bq
 {\bfJ}_\rho =-  L^{\rho\rho}\!\cdot \nabla H_\rho - L^{\rho\bfm}\!:\!\nabla  H_\bfm
   - L^{\rho\si} \!\cdot  \nabla H_\si\,, 
   \label{Jrho0}
   \eq
where the 2-tensors $L^{\rho\rho}$ and  $L^{\rho\si}$ and the 3-tensor $L^{\rho\bfm}$ are contracted as in \eqref{bep}. 
From the  functional derivative  $H_\rho$  of \eqref{HdivNSF} and the local thermodynamic identities \eqref{thermo1},  we find
\bq
\nabla H_\rho = -\frac{\si}{\rho}\nabla T + \frac{1}{\rho}\nabla p -  (\nabla\bfv)\cdot \bfv 
\label{nabHrho}
\eq
and 
\bq
\nabla p  = \rho\nabla H_\rho  + (\nabla H_\bfm)\cdot \bfm + \si \nabla H_\si\,.
\label{nablap}
\eq
Thus the difference velocity $\bfw$ of \eqref{J_gen} can be written  as the following  linear combination  of $\na H_\rho$, $\na H_\bfm$ and $\na H_\si$:
\bq
    \bfw = \tilde{D} \rho\, \nabla H_\rho + \tilde{D}\, (\nabla H_\bfm)\cdot \bfm + \tilde{D} \hat{\si} \, \nabla H_\si\,, 
    \label{J_lin}
\eq
where we defined $\hat{\si} \coloneqq \si -\ga$.  Therefore, according to  \eqref{mass_flux_BNSF}
\bq
  \bfJ_\rho = - \tilde{D} \rho^2\, \nabla H_\rho - \tilde{D}\rho \, (\nabla H_\bfm)\cdot \bfm   - \tilde{D}\rho\hat{\si} \,  \nabla H_\si\,,
\eq
and comparison with \eqref{Jrho0} yields
\bq
L^{\rho\rho}\!=\tilde{D} \rho^2\, \bar{I}\,,\quad L^{\rho\bfm}\! = \tilde{D}\rho\, \bar{I}  \otimes \bfm\,, \quad  L^{\rho\si}\!=  \tilde{D}\rho\hat{\si}\,\bar{I}\,.
\eq
Similarly,  using \eqref{genNSFflux}, \eqref{mom_flux_BNSF},  and \eqref{J_lin}, 
 \bal
     \bar{J}_{\mathbf{m}} &=-  L^{\bfm\rho}\otimes\nabla H_\rho - L^{\bfm\bfm}\!:\!\nabla H_\bfm - L^{\bfm\si}\otimes\nabla H_\si 
      \nn\\
     &= -\4tensLa\!:\!\nabla H_{\bfm} - \bfm\, \otimes 
     \nn\\
     &\hspace{.5cm} \big( \tilde{D} \rho\, \nabla H_\rho + \tilde{D}\, (\nabla H_\bfm)\cdot \bfm + \tilde{D} \hat{\si} \, \nabla H_\si
     \big)\,;
\eal
whence we see
\bal
L^{\bfm\rho}&=  \tilde{D} \rho \, \bfm \,,\quad L^{\bfm\si}=  \tilde{D} \hat{\si}\, \bfm \,, \quad \mathrm{and}
\nn\\
L^{\bfm\bfm}&=  \4tensLa  + \tilde{D}\, \bfm\otimes  \bar{I} \otimes \bfm\,.
\eal
Note, using  our convention $(\4tensLa:\!\nabla \bfv)_{ij}=\La_{ijkl} \p_k v_l$ we have 
\bal
\big(  \bfm\otimes  \bar{I} \otimes \bfm):\nabla H_\bfm\big)_{ij}
&= ( m_i \de_{jk} m_l) \p_k v_l 
\nn\\
&=  m_i  m_l \p_j v_l\,.
\eal
Finally,  using \eqref{genNSFflux}, \eqref{entropy_flux_BNSF},  and \eqref{J_lin}, 
\bal
 {\bfJ}_s &= -    L^{\si\rho}\!\cdot \nabla H_\rho
 -  L^{\si\bfm}\!:\! \nabla H_\bfm   - L^{\si\si}\!\cdot \nabla H_\si 
      \nn\\
     &= - \frac{\tenska}{T}\cdot \nabla H_\si  
      \label{BNSFfluxJs0}\\
     &\hspace{.75cm} - \hat{\si}\big(
     \tilde{D} \rho\, \nabla H_\rho + \tilde{D}\, (\nabla H_\bfm)\cdot \bfm + \tilde{D} \hat{\si} \, \nabla H_\si \big)
    \nn ; 
     \eal
 whence we see
 \bal
 L^{\si\rho}&=   \tilde{D} \rho\hat{\si}\,  \bar{I}\,,\quad  L^{\si\si}=  \frac{\tenska}{T}  +  \tilde{D}  \hat{\si}^2\, \bar{I}
 \\
 &  L^{\si\bfm} = \tilde{D}\hat{\si}\,  \bar{I}\otimes \bfm \,.
 \eal

With these phenomenological coefficients  we obtain directly the operators $M$ and $\Si$.  Again $M$ is chosen as in \eqref{UTANSFM}, while $\Si$ is given  as follows:
\bal
\Si(dF,dG) &= \frac{1}{T}\Big[{\tilde{D} \rho^2}  \,  \nabla{F_\rho} \cdot  \nabla G_\rho\nn\\
&\hspace{-.5cm} + {\tilde{D}\rho\,\big(  \nabla F_\rho \cdot (\nabla G_\bfm)\cdot \bfm} + \nabla G_\rho  \cdot(\nabla F_\bfm)\cdot\bfm\big)\nn\\
&\hspace{-.5cm}+ {\tilde{D}\rho\hat{\si}}\,\big( \nabla F_\rho\cdot    \nabla G_\si + \nabla G_\rho\cdot     \nabla F_\si \big)\nn\\
&\hspace{-.5cm}+  \nabla F_\bfm : \big( { \4tensLa }   + {\tilde{D}} \, \bfm\otimes  \bar{I} \otimes \bfm
\big): \nabla G_\bfm \nn\\
&\hspace{-.5cm} + \tilde{D} \hat{\si} \,\big( \nabla F_\si \cdot  (\nabla G_\bfm) \cdot\bfm+\nabla G_\si  \cdot   (\nabla F_\bfm)  \cdot \bfm
 \big)\nn\\
&\hspace{-.5cm}+  \na F_\si \cdot  \left(\frac{\tenska}{T} +  {\tilde{D}\hat{\si}^2}\bar{I}\right)\cdot \nabla G_\si \Big]
\label{BigSig}\,.
\eal
Note in the penultimate line of \eqref{BigSig} we have used
\bq
\nabla F_\si \cdot \bar{I}\otimes \bfm :\!\nabla G_\bfm= \nabla F_\si \cdot  (\nabla G_\bfm) \cdot\bfm\,.
\eq
(Recall \eqref{bep}.) The metriplectic 4-bracket $(\,.\,,\,.\,;\,.\,,\,.\,)$  that comes from the K-N product of $M$ and $\Si$ is the following:
\begin{widetext}   
\bal
(F, K;G , N)  &=
\int_{\Om} \frac{K_\si\,N_\si}{T}\left[\tilde{D} \rho^2 \nabla F_\rho \cdot \na G_\rho\right. +\tilde{D} \rho\big(\na F_\rho \cdot\left(\na G_{\bfm}\right) \cdot \mathbf{m}+\nabla G_\rho \cdot\left(\nabla F_{\bfm}\right) \cdot \bfm\big)
\nn \\
& \quadd +\tilde{D} \rho \hat{\sigma}\left(\nabla F_\rho \cdot \nabla G_\sigma+\nabla G_\rho \cdot \nabla F_\sigma\right)  
  +\nabla F_{\bfm}:   ( \4tensLa 
  +\tilde{D} \bfm \otimes \bar{I} \otimes \bfm): \nabla G_{\bfm}
  \nn \\
& \qquad \qquad +\tilde{D} \hat{\sigma}\, \big(\nabla F_\sigma \cdot\left(\nabla G_{\bfm}\right) \cdot \bfm+\nabla G_\sigma \cdot\left(\nabla F_{\bfm}\right) \cdot \bfm\big)  \left.+\nabla F_\sigma \cdot\left(\frac{\bar{\kappa}}{T}+\tilde{D} \hat{\sigma}^2 \bar{I}\right) \cdot \nabla G_\sigma\right]
\nn\\
& - \frac{K_\si\,G_\si}{T}\left[\tilde{D} \rho^2 \nabla F_\rho \cdot \na N_\rho\right.  +\tilde{D} \rho\, \big(\na F_\rho \cdot\left(\na N_{\bfm}\right) \cdot \mathbf{m}+\nabla N_\rho \cdot\left(\nabla F_{\bfm}\right) \cdot \bfm\big)
\nn \\
&\qquad  +\tilde{D} \rho \hat{\sigma}\left(\nabla F_\rho \cdot \nabla N_\sigma+\nabla N_\rho \cdot \nabla F_\sigma\right)  +\nabla F_{\bfm}:(\4tensLa +\tilde{D} \bfm \otimes \bar{I} \otimes \bfm): \nabla N_{\bfm}
\nn \\
& \qquad \qquad +\tilde{D} \hat{\sigma}\left(\nabla F_\sigma \cdot\left(\nabla N_{\bfm}\right) \cdot \bfm+\nabla N_\sigma \cdot\left(\nabla F_{\bfm}\right) \cdot \bfm\right)  
 \left.+\nabla F_\sigma \cdot\left(\frac{\bar{\kappa}}{T}+\tilde{D} \hat{\sigma}^2 \bar{I}\right) \cdot \nabla N_\sigma\right]\nn\\
& + \frac{F_\si\,G_\si}{T}\left[\tilde{D} \rho^2 \nabla K_\rho \cdot \na N_\rho\right. 
  +\tilde{D} \rho\, \big(\na K_\rho \cdot\left(\na N_{\bfm}\right) \cdot \mathbf{m}+\nabla N_\rho \cdot\left(\nabla K_{\bfm}\right) \cdot \bfm\big) 
  \nn\\
& \qquad +\tilde{D} \rho \hat{\sigma}\left(\nabla K_\rho \cdot \nabla N_\sigma+\nabla N_\rho \cdot \nabla K_\sigma\right)  
  +\nabla K_{\bfm}:(\4tensLa +\tilde{D} \bfm \otimes \bar{I} \otimes \bfm): \nabla N_{\bfm} 
  \nn\\
&\qquad  \qquad +\tilde{D} \hat{\sigma} \left(\nabla K_\sigma \cdot\left(\nabla N_{\bfm}\right) \cdot \bfm+\nabla N_\sigma \cdot\left(\nabla K_{\bfm}\right) \cdot \bfm\right)  
 \left.+\nabla K_\sigma \cdot\left(\frac{\bar{\kappa}}{T}+\tilde{D} \hat{\sigma}^2 \bar{I}\right) \cdot \nabla N_\sigma\right]
\nn\\
&- \frac{F_\si\,N_\si}{T}\left[\tilde{D} \rho^2 \nabla K_\rho \cdot \na G_\rho\right.  
 +\tilde{D} \rho\, \big(\na K_\rho \cdot\left(\na G_{\bfm}\right) \cdot \mathbf{m}+\nabla G_\rho \cdot\left(\nabla K_{\bfm}\right) \cdot \bfm\big) 
 \nn\\
& \qquad +\tilde{D} \rho \hat{\sigma}\left(\nabla K_\rho \cdot \nabla G_\sigma+\nabla G_\rho \cdot \nabla K_\sigma\right)  +\nabla K_{\bfm}:(  \4tensLa +\tilde{D} \bfm \otimes \bar{I} \otimes \bfm): \nabla G_{\bfm} 
\nn\\
&\qquad \qquad +\tilde{D} \hat{\sigma}\, \big(\nabla K_\sigma \cdot\left(\nabla G_{\bfm}\right) \cdot \bfm+\nabla G_\sigma \cdot\left(\nabla K_{\bfm}\right) \cdot \bfm\big)  \left.+\nabla K_\sigma \cdot\left(\frac{\bar{\kappa}}{T}+\tilde{D} \hat{\sigma}^2 \bar{I}\right) \cdot \nabla G_\sigma\right].
\eal
\end{widetext}
Upon insertion of  $S$ as given by \eqref{S_nsf}  and $H$ as given by \eqref{Ham_1}, the system \eqref{rho_BNSF}, \eqref{m_BNSF}, and \eqref{sigma_BNSF} is produced according to 
     \bal
     \p_t\rho &= \{\rho, H\} + (\rho,H;S,H)\,, 
\nn\\
     \p_t\bfm &= \{\bfm, H\} + (\bfm,H;S,H)\,, 
 \nn\\
     \p_t\si &= \{\si, H\} + (\si,H;S,H)\,,
     \nn
    \eal
and the total entropy production  is  governed by the following:
\bal
\dot{S}& = (S,H;,S,H) = \int_\Om \Si(dH,dH)\,\nn\\
&= \int_\Om \frac{1}{\tilde{D}\,T}\bfw\cdot\bfw  +\nabla T\cdot\frac{\tenska}{T}\cdot\nabla T+\nabla \bfv : \4tensLa:\nabla  \bfv\,\nn\\
&= \int_\Om \frac{1}{T}\left[{\tilde{D}}|\bfv_m - \bfv|^2  +\nabla T\cdot\frac{\tenska}{T}\cdot\nabla T\right.\nn\\
&\hspace{4cm}\left.+\nabla \bfv : \4tensLa:\nabla  \bfv\right] \geq 0 \,.
\eal
Alternatively, using \eqref{BrkaT}
\bal
\dot{S}=\int_\Om \frac{1}{T}&\left[\frac{\tilde{D}}{\kappa_T^2 \rho^2}|\nabla \rho|^2+\nabla T\cdot\frac{\tenska}{T}\cdot\nabla T\right.\nn\\
&\hspace{2cm}\left. +\nabla \bfv : \4tensLa:\nabla  \bfv\right]\geq 0 \,.
\eal

Therefore, we shown  that the system proposed by Brenner \cite{BRENNER2006} can be understood as an extension of the classical Navier-Stokes-Fourier, achieved by introducing an additional dissipation mechanism. Brenner postulates that his hypothesis primarily alters the ideal part of the dynamics.  However, if by ideal is meant Hamiltonian, we see that this  is not true since the Hamiltonian part is still governed by the Poisson bracket of \cite{pjmG80}.   In addition, Brenner  links this modification to the compressibility of the fluid and   suggests that the mass velocity $\bfv_m$ and volume velocity $\bfv$ coincide if, and only if, the fluid is incompressible (i.e., $\rho = const$).

We have also shown that  the expression of $\bfw$ given by \eqref{J_gen} is not  the most general form giving a thermodynamically consistent system; from \eqref{Jrho0} 
\bq
 {\bfw} = \big(L^{\rho\rho}\!\cdot \nabla H_\rho + L^{\rho\bfm}\!:\!\nabla  H_\bfm + 
    L^{\rho\si} \!\cdot  \nabla H_\si\big)/\rho\,, 
   \label{Bgeneralization}
   \eq
 whence we see that $\bfw$ can be any  linear combination  of $\na H_\rho$, $\na H_\bfm$ and $\na H_\si$ contracted appropriately with  the 2-tensors $L^{\rho\rho}$ and  $L^{\rho\si}$ and the 3-tensor $L^{\rho\bfm}$.

In a more recent paper \cite{Reddy19}  thermodynamically consistent generalizations of the BNSF system were given.  
In concluding this section we show that the various generalizations of this reference are again special cases of our metriplectic system of Sec.~\ref{NSF} with \eqref{Bgeneralization}.  Specifically,  the cases of  \cite{Reddy19}   (rewritten in our notation)  are as follows:

\medskip
\noindent
Equation (77) of   \cite{Reddy19}, 
\begin{equation} 
\bfw=   \kappa_m \nabla \ln{\rho}\,,   
\eq
where $\ka_m = \tilde{D}/\kappa_T$, is Brenner's  \eqref{BrkaT} using $\gamma = \left(\frac{\partial p}{\partial T}\right)_\rho$; 
\\
Equation (78) of  \cite{Reddy19}, 
\begin{equation}
\bfw=\kappa_T \nabla \ln{T} =   \frac{\kappa_T}{T} \nabla H_{\sigma}\,, 
\label{78}
\end{equation}
is given by  our  \eqref{Bgeneralization} with the choices 
\bq
L^{\rho\rho} = L^{\rho\mathbf{m}} = 0,\quad L^{\rho\sigma} =\rho \, \frac{\kappa_T}{T}\, \bar{I} \,;  
\eq
\\
Equation (79) of  \cite{Reddy19} is, 
\bal
\bfw &= \kappa_p \nabla \ln{p} =   \frac{\kappa_p}{p} \nabla p
\nn\\
& =  \frac{\kappa_p}{p} \big(\rho\nabla H_\rho  +   (\nabla H_\mathbf{m})\cdot \bfm + \sigma \nabla H_\sigma\big)\,,
\label{79}
\eal
where $\ka_p$ is the thermal conductivity at constant pressure and the third  equality follows from \eqref{nablap}.  Equation \eqref{79} is given by our   \eqref{Bgeneralization} with the choices 
\bal
L^{\rho\rho} &= \rho^2 \frac{\kappa_p}{p}\,\bar{I}\,, \quad  L^{\rho\mathbf{m}} =  {\rho\,  \kappa_p} \,\bar{I} \otimes \mathbf{m}\,;
\nn\\
 L^{\rho\sigma} &= \rho \,  {\kappa_p\,\sigma}\,\bar{I}\,.
\eal
\\
Equation (80) of  \cite{Reddy19} is, 
\begin{equation}
\bfw=  \kappa_{\tau} \nabla \times {\bfv} = {\kappa_{\tau}} \nabla \times H_{\mathbf{m}} \,,
\label{80}
\end{equation}
 where $\ka_\ta$ is another phenomenological quantity.  Equation  \eqref{80}   is a particular case of our theory by taking
\bq
 L^{\rho\rho} = 0,\quad  L^{\rho\mathbf{m}} = \rho \, {\kappa_{\tau}} \, \bm{\ep}\,, 
 \quad L^{\rho\sigma} = 0\,, 
 \eq
where $\bm{\ep}$ is the Levi-Civita 3-tensor  (density) and contraction is defined by \eqref{bep}.  Note,  the tensorial inconsistency of \eqref{80} can be resolved by assuming  $\ka_T$ is a pseudoscalar.

\section{Conclusion}
\label{conclusion}

The  main contribution of this paper is the unified thermodynamical algorithm  that uses the metriplectic 4-bracket of previous work \cite{pjmU24,zaidni2,pjmS24} to methodically lead one to general classes of thermodynamically consistent  systems.  An important and novel by-product of this algorithm is the definition of fluxes given by \eqref{genOnsag}.  In Sec.~\ref{Overview on metriplectic framework} we reviewed the Hamiltonian and 4-bracket frameworks, on which the UT algorithm is based.  This is followed by  Sec.~\ref{Derivation of metriplectic 4-bracket} that contains  the main new contribution:  the unambiguous determination of the metriplectic 4-bracket.  In Sec.~\ref{Examples} we present examples that generalize previous results.  In particular,  we showed that the Brenner-Navier-Stokes-Fourier system and its generalizations of \cite{Reddy19} are  special cases of our generalization of the Navier-Stokes-Fourier system.  They all amount to modifying the dissipation in the Navier-Stokes equations. 

The dichotomies of  dissipative vs.\  nondissipative and reversible vs.\  irreversible can be  confused or used inappropriately,  particularly when one is dealing with  systems that contain  a set of conservation laws such as those  of \eqref{N-1-evol}.  One clear distinction can be made:  that between  Hamiltonian vs.\  nonHamiltonian, where the former is  an unambiguous  definition of what is meant by nondissipative.  The distinction between reversible and irreversible is also often confused.  All systems of autonomous ordinary differential equations are reversible because the solution is a one-parameter Lie group, and not all Hamiltonian systems have time reversible symmetry, a special case of a point symmetry.  Again, there is no confusion if one distinguishes Hamiltonian from nonHamiltonian, and the metriplectic 4-bracket formalism makes it  clear which parts are  Hamiltonian and which parts are dissipative. 

Another dichotomy  concerns the placement of temperature in the metriplectic formalism.    Temperature may appear as a result of the assumption of  local thermodynamic equilibrium, e.g., via an internal energy function $u$ in the Hamiltonian,  or it may appear in the assumed forms of the phenomenological coefficients $L^{\al\be}$.  In the first  work on the metriplectic dynamics of the NSF fluid \cite{pjm84b}, it  was observed  that the temperature needed to be placed in an ad hoc manner  so as to make things work out.  Similarly,  the same observation  was  noted  in Chap.\ 3 of  \cite{ottinger2005}.   A resolution of this dichotomy is achieved with the UT algorithm, where temperature may appear according to  \eqref{N-1-evol} and \eqref{UTASi} or in the choice of phenomenological coefficients.  It is interesting to note that once $M$ and  $\Si$ are chosen and the 4-bracket is determined, one can use any Hamiltonian and obtain a thermodynamically  consistent system. This provides additional freedom for modeling.

In closing we mention some possibilities for future work.  The results of this paper pertain to  macroscopic or purely continuum theories.  Underlying kinetic theory can place constraints on such continuum theories.  For example, in  \cite{Mills06} it was noted that the results of Brenner are in disagreement with a number of kinetic-theory studies.   In the present context, an open question is how  to connect the 4-bracket to a class of underlying kinetic theories with dissipative mechanisms such as collision operators.  On the kinetic level, a  metriplectic 4-bracket was given in \cite{pjmS24} for a generalization of the Landau collision operator  and the same can be done for a variety of kinetic theories.    So far, no connection has been made between fluid and kinetic 4-brackets.

The UT algorithm can be both restricted and generalized.  For example, additional symmetries beyond Onsager,  such as Galilean  or Poincar\'e invariance,  can constrain the choices of $M$ and  $\Si$ .  These symmetries might be traced from a kinetic theory or considered on the macroscopic  level. Here we have not considered these possibilities, so as to keep the development  general.  An avenue for further generalization would be to break the  linear force-flux relations of \eqref{onsager1} or  \eqref{genOnsag}.  The essential feature of thermodynamic consistency is global asymptotic stability and the concomitant production of entropy.  Dynamical systems with global asymptotic stability   can be recast into the form of \eqref{onsager1} or  \eqref{genOnsag} by using rectification arguments similar to those described in  \cite{pjmU24}. Rectification arguments fail when additional fixed points exist.  Systems with this property would not be expected to be thermodynamically consistent, but one could  still linearize within basins of attraction.

{\section*{Acknowledgements}

A.Z. acknowledges support from the Mohammed VI Polytechnic University for supporting an internship at the University of Texas at Austin.  He would also like to thank  R.\ Boukharfane for his support.  P.J.M. acknowledges support from the DOE Office of Fusion Energy Sciences under DE-FG02-04ER-54742, and  would like to thank William Barham and Chris Eldred for helpful conversations.}








%
\bibliographystyle{unsrt}


%


\end{document}